\documentstyle[times,psfig]{mn}

\newif\ifAMStwofonts

\ifoldfss
  \newcommand{\rmn}[1] {{\rm #1}}

  \ifCUPmtlplainloaded \else
    \NewTextAlphabet{textbfit} {cmbxti10} {}
    \NewTextAlphabet{textbfss} {cmssbx10} {}
    \NewMathAlphabet{mathbfit} {cmbxti10} {} 
    \NewMathAlphabet{mathbfss} {cmssbx10} {} 
  \fi
  \ifAMStwofonts
    \ifCUPmtlplainloaded \else
      \NewSymbolFont{upmath} {eurm10}
      \NewSymbolFont{AMSa} {msam10}
      \NewMathSymbol{\upi}     {0}{upmath}{19}
      \NewMathSymbol{\umu}     {0}{upmath}{16}
      \NewMathSymbol{\upartial}{0}{upmath}{40}
      \NewMathSymbol{\leqslant}{3}{AMSa}{36}
      \NewMathSymbol{\geqslant}{3}{AMSa}{3E}

       \let\le=\leqslant
       \let\ge=\geqslant
    \fi
  \fi
\fi 

\ifnfssone
  \newmathalphabet{\mathit}
  \addtoversion{normal}{\mathit}{cmr}{m}{it}
  \addtoversion{bold}{\mathit}{cmr}{bx}{it}
  \newcommand{\rmn}[1] {\mathrm{#1}}

  \newmathalphabet{\mathbfit} 
  \addtoversion{normal}{\mathbfit}{cmr}{bx}{it}
  \addtoversion{bold}{\mathbfit}{cmr}{bx}{it}
  \newmathalphabet{\mathbfss} 
  \addtoversion{normal}{\mathbfss}{cmss}{bx}{n}
  \addtoversion{bold}{\mathbfss}{cmss}{bx}{n}
  \ifAMStwofonts
    \ifCUPmtlplainloaded \else
      \UseAMStwoboldmath
      \makeatletter
      \new@mathgroup\upmath@group
      \define@mathgroup\mv@normal\upmath@group{eur}{m}{n}
      \define@mathgroup\mv@bold\upmath@group{eur}{b}{n}
      \edef\UPM{\hexnumber\upmath@group}
      \new@mathgroup\amsa@group
      \define@mathgroup\mv@normal\amsa@group{msa}{m}{n}
      \define@mathgroup\mv@bold\amsa@group{msa}{m}{n}
      \edef\AMSa{\hexnumber\amsa@group}
      \makeatother
      \mathchardef\upi="0\UPM19
      \mathchardef\umu="0\UPM16
      \mathchardef\upartial="0\UPM40
      \mathchardef\leqslant="3\AMSa36
      \mathchardef\geqslant="3\AMSa3E

       \let\le=\leqslant
       \let\ge=\geqslant
    \fi
  \fi
\fi 

\ifnfsstwo
  \newcommand{\rmn}[1] {\mathrm{#1}}

  \DeclareMathAlphabet{\mathbfit}{OT1}{cmr}{bx}{it}
  \SetMathAlphabet\mathbfit{bold}{OT1}{cmr}{bx}{it}
  \DeclareMathAlphabet{\mathbfss}{OT1}{cmss}{bx}{n}
  \SetMathAlphabet\mathbfss{bold}{OT1}{cmss}{bx}{n}
  \ifAMStwofonts
    \ifCUPmtlplainloaded \else
      \DeclareSymbolFont{UPM}{U}{eur}{m}{n}
      \SetSymbolFont{UPM}{bold}{U}{eur}{b}{n}
      \DeclareSymbolFont{AMSa}{U}{msa}{m}{n}
      \DeclareMathSymbol{\upi}{0}{UPM}{"19}
      \DeclareMathSymbol{\umu}{0}{UPM}{"16}
      \DeclareMathSymbol{\upartial}{0}{UPM}{"40}
      \DeclareMathSymbol{\leqslant}{3}{AMSa}{"36}
      \DeclareMathSymbol{\geqslant}{3}{AMSa}{"3E}

       \let\le=\leqslant
       \let\ge=\geqslant
    \fi
  \fi
\fi 

\ifCUPmtlplainloaded \else
  \ifAMStwofonts \else 
    \def\upi{\pi}
    \def\umu{\mu}
    \def\upartial{\partial}
  \fi
\fi

   \title[Deuterium astration]{Deuterium astration in the local disc and 
                               beyond}
   \author[D. Romano et al.]{Donatella Romano,$^{1}$\thanks{E-mail: 
           donatella.romano@oabo.inaf.it (DR); monica.tosi@oabo.inaf.it (MT); 
           chiappini@ts.astro.it (CC); matteucci@ts.astro.it (FM)} Monica 
           Tosi,$^{1}$ Cristina Chiappini$^{2, 3}$ and Francesca 
	   Matteucci$^{4}$\\
           $^{1}$INAF\,--\,Osservatorio Astronomico di Bologna,
                 Via Ranzani 1, I-40127 Bologna, Italy\\
           $^{2}$INAF\,--\,Osservatorio Astronomico di Trieste,
                 Via Tiepolo 11, I-34131 Trieste, Italy\\
           $^{3}$Geneva Observatory,  
                 1290 Sauverny, Switzerland\\
           $^{4}$Dipartimento di Astronomia, Universit\`a di Trieste,
                 Via Tiepolo 11, I-34131 Trieste, Italy}
   
\begin{document}

     \date{Accepted 2006 March 7. Received 2006 March 7; in original form 2006 
           January 19}

     \pagerange{\pageref{firstpage}--\pageref{lastpage}} \pubyear{2006}

     \maketitle

     \label{firstpage}


   \begin{abstract}
     Estimates of the interstellar deuterium abundance span a wide range of 
     values. Until recently, it was customary to adopt the abundance of 
     deuterium measured in the Local Bubble as representative of the local 
     one. Now, it is becoming unclear whether the true local deuterium 
     abundance is significantly higher or lower than this value, depending on 
     the interpretation given to current data. It is important to deal with 
     the issue of the deuterium variation and see whether it challenges our 
     current understanding of the Galaxy evolution. To this aim, we study the 
     evolution of deuterium in the framework of successful models for the 
     chemical evolution of the Milky Way able to reproduce the majority of the 
     observational constraints for the solar neighbourhood and for the 
     Galactic disc. We show that, in the framework of our models, the lowest 
     D/H values observed locally cannot be explained in terms of simple 
     astration processes occurring during the Galaxy evolution. Indeed, the 
     combination of a mild star formation and a continuous infall of 
     unprocessed gas required to fit all the available observational data 
     allows only a modest variation of the deuterium abundance from its 
     primordial value. Therefore, we suggest that depletion of deuterium on to 
     dust grains is the most likely physical mechanism proposed so far to 
     explain the observed dispersion in the local data.
   \end{abstract}

   \begin{keywords}
     cosmology: observations -- Galaxy: abundances -- Galaxy: evolution -- 
     ISM: abundances
   \end{keywords}


   \section{Introduction}
   \label{sec.int}

   Deuterium is created during primordial nucleosynthesis (e.g. Weinberg 1977; 
   Boesgaard \& Steigman 1985; Olive, Steigman \& Walker 2000) and then 
   destroyed in stars through D\,(p, $\gamma$)\,$^3$He reactions occurring at 
   relatively low temperatures (of a few 10$^5$ K). Since it is hard to 
   synthesize this loosely bound isotope in significant quantities in any 
   astrophysical environment (Epstein, Arnett \& Schramm 1976; Epstein 1977; 
   Prodanovi\'c \& Fields 2003), its abundance in galaxies is expected to 
   monotonically decrease with time owing to stellar cycling.

   Deuterium is the light nuclide produced during Big Bang Nucleosynthesis 
   (BBN) whose primordial abundance is most tightly tied to the 
   baryon-to-photon ratio, $\eta$ (or, equivalently, to the ratio of the 
   baryon and critical densities at the present time, $\Omega_{\mathrm{b}}$). 
   In principle, once its primordial abundance, (D/H)$_{\mathrm{p}}$, is 
   derived from measurements in unevolved systems, the value of 
   $\Omega_{\mathrm{b}}$ is precisely known. However, there is still 
   significant dispersion among the abundances derived for a handful of 
   high-redshift, low-metallicity absorption-line systems towards 
   quasi-stellar objects (QSOs) with reasonably firm deuterium detections. A 
   more precise $\eta$ determination comes from the observations of the cosmic 
   microwave background (CMB) power spectrum coupled to Ly$\alpha$ forest data 
   (Spergel et al. 2003; see, however, Pettini 2006, his fig.~7). The 
   suggested value, $\eta_{10}~= 6.1^{+0.3}_{-0.2}$, where $\eta_{10} \equiv 
   10^{10} \eta = 274 \, \Omega_{\mathrm{b}} h^2$, when coupled to predictions 
   from standard BBN theory, leads to a very narrow range for the primordial 
   deuterium abundance, (D/H)$_{\mathrm{p}} \simeq$ 
   2.4--2.8~$\times$~10$^{-5}$ (e.g. Coc et al. 2004).

   Relating (D/H)$_{\mathrm{p}}$ to the current local abundance of D meets 
   with the non-trivial issue of estimating what the \emph{true} local 
   abundance of deuterium actually is. In fact, a large scatter is found among 
   nearly fifty D/H measurements towards as many Galactic lines of sight. 
   Possible explanations (including deuterium depletion on to dust grains, or 
   recent infall of unprocessed gas; see e.g. Pettini 2006, and references 
   therein; Prochaska et al. 2005) lead to different interpretations of the 
   data.

%
   \begin{table*}
   \caption[]{D/H measurements towards QSOs. A weighted mean of these data 
              (except measurements towards Q\,0347$-$3819 and PKS\,1937$-$1009 
              at $z_{\rmn{abs}}$ = 3.256; see text for details) gives 
	      $\langle$D/H$\rangle$ = 2.6 $\pm$ 0.3 (the error in the mean is 
              the dispersion about the mean divided by the square root of the 
	      number of data points), a value consistent with the primordial 
	      abundance inferred from the CMB power spectrum and the standard 
	      BBN theory.
             }
   \begin{center}
   \begin{tabular}{l c c c c c c l}
   \hline
   QSO & Target$^{a}$ & log\,$N$(H\,{\small I}) & $z_{\rmn{abs}}$ & D/H $\pm 1 \sigma$ & [Si/H] & [O/H] & Reference\\
   \hline
   PKS\,1937$-$1009\,(I) & LL & 17.86 $\pm$ 0.02 & 3.572 & (3.3 $\pm$ 0.3) $\times 10^{-5}$ & $-$2.7, $-$1.9$^{b}$ & & Burles \& Tytler 1998a\\
   Q\,1009+2956 & LL & 17.39 $\pm$ 0.06 & 2.504 & 3.98$^{+0.59}_{-0.67}$ $\times 10^{-5}$ & $-$2.4, $-$2.7$^{c}$ & & Burles \& Tytler 1998b\\
   HS\,0105+1619 & LL & 19.422 $\pm$ 0.009 & 2.536 & (2.54 $\pm$ 0.23) $\times 10^{-5}$ & $-$1.85 & $-$2.0 & O'Meara et al. 2001\\
   Q\,2206$-$199 & DLA & 20.436 $\pm$ 0.008 & 2.0762 & (1.65 $\pm$ 0.35) $\times 10^{-5}$ & $-$2.23 & & Pettini \& Bowen 2001\\
   Q\,0347$-$3819 & DLA & 20.626 $\pm$ 0.005 & 3.025 & (3.75 $\pm$ 0.25) $\times 10^{-5}$ & $-$0.95 $\pm$ 0.02 & & Levshakov et al. 2002\\
   Q\,1243+3047 & LL & 19.76 $\pm$ 0.04 & 2.526 & 2.42$^{+0.35}_{-0.25}$ $\times 10^{-5}$ & & $-$2.79 $\pm$ 0.05 & Kirkman et al. 2003\\
   PKS\,1937$-$1009\,(II) & LL & 18.25 $\pm$ 0.02 & 3.256 & 1.6$^{+0.25}_{-0.30}$ $\times 10^{-5}$ & $-$2.0 $\pm$ 0.5 & & Crighton et al. 2004\\
   PG\,1259+593 & HVC & 19.95 $\pm$ 0.05 & 0.0 & (2.2 $\pm$ 0.7) $\times 10^{-5}$ & & $-$0.79$^{+0.12}_{-0.16}$ & Sembach et al. 2004\\
   \hline
   \end{tabular}
   \end{center}
\begin{list}{}{}
\item[$^{a}$]LL = Lyman Limit absorption system; DLA = Damped Ly$\alpha$ 
             absorption system; HVC = high-velocity cloud.
\item[$^{b}$]Blue component, red component.
\item[$^{c}$]Component 2, component 3.
\end{list}
   \label{tab.qso}
   \end{table*}
%

   Up to some years ago, models for the chemical evolution of the Galactic 
   disc were used to bound the primordial D mass fraction. Studies of this 
   kind were motivated by the widely differing (by an order of magnitude) 
   observational estimates of (D/H)$_{\mathrm{p}}$ published in the 
   literature, coming from a few high-redshift QSO absorption spectra. It was 
   crucial then to infer (D/H)$_{\mathrm{p}}$ from the D abundance observed in 
   the solar system and the interstellar medium (ISM) assuming the astration 
   factor required to reproduce the largest set of available observational 
   data. In general, it was found that (D/H)$_{\mathrm{p}}$ values higher than 
   $\sim$4--5~$\times$10$^{-5}$ could not be reconciled with Galactic chemical 
   evolution (GCE) requirements (Steigman \& Tosi 1992, 1995; Tosi 1996, and 
   references therein; Tosi et al. 1998; Prantzos \& Ishimaru 2001; Chiappini, 
   Renda \& Matteucci 2002), with sparse claims of the contrary (e.g. 
   Vangioni-Flam, Olive \& Prantzos 1994; Scully et al. 1997).

   Recently, Kirkman et al. (2001) obtained Space Telescope Imaging 
   Spectrograph (STIS) spectra of the Ly$\alpha$ and Lyman Limit (LL) regions 
   of the $z_{\rmn{abs}}$~= 0.701 absorption system towards QSO PG\,1718+4807 
   (the only example of a QSO absorber with a D/H ratio $\sim$10 times the 
   value found towards other QSOs). They found that at least part of the extra 
   absorption needed on the blue side of the main H\,{\small I}, which was 
   previously ascribed to deuterium, is more likely due to contaminating 
   hydrogen (see, however, Crighton et al. 2003). Even more recently, a few 
   more low-metallicity, high-redshift QSO absorption systems have been 
   searched for deuterium, and all show low D abundances (see next section). 
   This piece of evidence, together with the concordance between the $\eta$ 
   value obtained from the temperature anisotropies in the CMB (as well as 
   other astronomical measurements of the power spectrum) and that suggested 
   by the standard BBN theory for low (D/H)$_{\mathrm{p}}$ values, leads to 
   the conclusion that (D/H)$_{\mathrm{p}}$ is indeed low, not in excess of 
   3~$\times$~10$^{-5}$ (see, e.g., discussions in Romano et al. 2003; and 
   recent reviews by Steigman 2004, 2006; and Pettini 2006). At this point, it 
   is worth mentioning that an internal tension between the abundances of the 
   light elements created in BBN might exist -- the primordial abundances of 
   $^4$He and $^7$Li inferred from the observations being lower than those 
   expected according to the BBN theory constrained by \emph{WMAP} (see, e.g., 
   fig.~5 of Pettini 2006). While for $^4$He one can appeal to previously 
   unrecognized systematic uncertainties in the measurements (Olive \& 
   Skillman 2004), dealing with $^7$Li in halo stars poses awkward 
   implications to stellar evolutionary theories. Yet, the inclusion of 
   non-standard depletion mechanisms in stellar evolutionary models promises 
   to reconcile the abundances observed in halo dwarfs with the primordial 
   $^7$Li abundance inferred from the CMB power spectrum and the standard BBN 
   theory (see Charbonnel \& Primas 2005 for a discussion).

   GCE models are however facing a new problem: connecting the reasonably 
   well-known, low primordial abundance of deuterium with a presently 
   uncertain local value. In fact, according to the newest data, the local 
   abundance of deuterium might be either nearly primordial, or significantly 
   lower, quite lower than previously assumed [(D/H)$_{\mathrm{LISM}}$~= (1.50 
   $\pm$ 0.10)~$\times$~10$^{-5}$; Linsky 1998], depending on the data 
   interpretation.

   In this paper we will examine these new local D data in the framework of 
   Galactic chemical evolution models able to reproduce the vast majority of 
   the available constraints on the disc properties. The paper is organized as 
   follows: measurements of deuterium at both high and low redshifts are 
   reviewed in Section~\ref{sec.dat}. The dispersion in the local data and its 
   implications are discussed in Section~\ref{sec.dev}, where the data are 
   interpreted in the framework of complete GCE models. Finally, a discussion 
   is presented in Section~\ref{sec.fin}.

   \section{Local and high-redshift deuterium data}
   \label{sec.dat}

\begin{figure}
   \psfig{figure=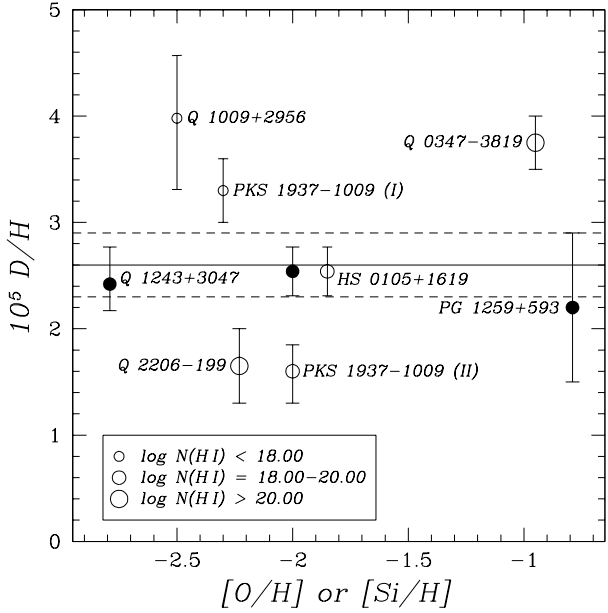,width=\columnwidth}
      \caption{ Available measurements of D/H in QSO absorption systems from 
	        Table~\ref{tab.qso}. The abundance of deuterium is plotted 
		against metallicity (either [Si/H] -- open symbols, or [O/H] 
		-- filled ones). There are no hints of decreasing D/H with 
		increasing metallicity. Rather, a large dispersion is present 
		across the whole metallicity range. The weighted mean of the 
		data (except for measurements towards Q\,0347$-$3819 and 
		PKS\,1937$-$1009 at $z_{\rmn{abs}}$ = 3.256; see text for 
		details) is also shown (solid line) with its errors (dashed 
		lines).
              }
         \label{fig.dsio}
   \end{figure}

   Data on high-redshift, low-metallicity Damped Lyman-$\alpha$ (DLA) and LL 
   absorption systems set lower limits to the primordial abundance of 
   deuterium. There is statistically significant scatter in the D/H 
   measurements for about half a dozen quasar absorption systems with 
   [Si/H]~$\la$~$-$1.0 about the mean (see Table~\ref{tab.qso}). The system at 
   $z_{\rmn{abs}}$~= 3.025 towards Q\,0347$-$3819 and that at 
   $z_{\rmn{abs}}$~= 3.256 towards PKS\,1937$-$1009 are excluded from our 
   analysis because of their complex velocity structures, which make them 
   rather unreliable (Steigman 2004, 2006; Pettini 2006). Since the primordial 
   D/H is thought to be isotropic and homogeneous and the metallicity of the 
   absorption clouds used to measure D/H is low (hence no significant D 
   astration through stars should have occurred), the scatter is hard to 
   explain. It has been proposed that some early mechanism for D astration may 
   be the cause for the reported scatter in D/H values (Fields et al. 2001; 
   Prantzos \& Ishimaru 2001). Yet, the correlation between D/H and [Si/H] 
   that one would expect in this case is not seen (Fig.~\ref{fig.dsio}). On 
   the other hand, unrecognized systematic effects may be responsible for an 
   artificial dispersion in the data. In this respect, it is interesting to 
   notice that the total neutral hydrogen column density $N$(H\,{\small I}) 
   for the two systems with the highest D/H values at [Si/H]~$\la$~$-$2.0 was 
   measured using the drop in flux at the Lyman limit. In all the other 
   systems, the dominant constraint on the $N$(H\,{\small I}) value is from 
   the damping wings of the Ly$\alpha$ or Ly$\beta$ lines (Pettini \& Bowen 
   2001; Crighton et al. 2004). Still, the placement of the continuum in the 
   systems with the highest $N$(H\,{\small I}) might be compromised by the 
   presence of interlopers, and lead to erroneous $N$(H\,{\small I}) 
   determinations (Steigman 1994, 2006).

%
%

   Further measurements of deuterium in metal-poor environments include that 
   in the high-velocity, low-metallicity gas cloud Complex C falling on to the 
   Milky Way, also listed in Table~\ref{tab.qso}. Using \emph{Far Ultraviolet 
   Spectroscopic Explorer (FUSE)} and \emph{Hubble Space Telescope (HST)} 
   observations of the QSO PG\,1259+593, Sembach et al. (2004) detect 
   D\,{\small I} Lyman series absorption in Complex C and derive 
   (D/H)$_{\rmn{Complex \ C}}$~= (2.2~$\pm$ 0.7)~$\times$ 10$^{-5}$, 
   (O/H)$_{\rmn{Complex \ C}}$~= (8.0~$\pm$ 2.5)~$\times$ 10$^{-5}$, and 
   (D/O)$_{\rmn{Complex \ C}}$~= 0.28~$\pm$ 0.12. A value consistent with the 
   primordial deuterium abundance, $\langle$D/H$\rangle$~= (2.3~$\pm$ 0.4) 
   $\times$ 10$^{-5}$ (weighted mean for three regions at Galactic longitudes 
   $l$ = 171$^\circ$, 183$^\circ$, 195$^\circ$), is found also from the 327 
   MHz D line, for a Galactocentric distance of $R_{\rmn{G}}$ = 10~$\pm$ 1 kpc 
   (Rogers et al. 2005).

   Precise measurements of D/H in the local interstellar medium (LISM) were 
   obtained with \emph{Copernicus} (e.g. Rogerson \& York 1973), \emph{HST} 
   (e.g. Linsky 1998), and the Interstellar Medium Absorption Profile 
   Spectrograph (IMAPS; Jenkins et al. 1999; Sonneborn et al. 2000). In recent 
   years, \emph{FUSE} has added many determinations of D/H (as well as D/O and 
   D/N) along several lines of sight probing the neutral ISM up to $\sim$2.7 
   kpc away (e.g. Moos et al. 2002; H\'ebrard \& Moos 2003; Wood et al. 2004; 
   H\'ebrard et al. 2005; Oliveira et al. 2006). As data accumulated, the 
   situation got complicated. Within the Local Bubble (LB; a low-density 
   region within a distance of $\sim$100 pc in which the Sun is embedded), D/H 
   is nearly constant. However, while Wood et al. (2004) state that 
   (D/H)$_{\rmn{LB}}$~= (1.56~$\pm$ 0.04)~$\times$ 10$^{-5}$, a value derived 
   from $N$(D\,{\small I}) and $N$(H\,{\small I}) measurements towards 16 
   targets, H\'ebrard \& Moos (2003), who rely on measurements of D/O and O/H, 
   find (D/H)$_{\rmn{LB}}$~= (1.32~$\pm$ 0.08)~$\times$ 10$^{-5}$, a value 
   significantly lower. At log\,$N$(H\,{\small I}) $\ga$ 20.5 (i.e., distances 
   greater than 500 pc), the few data points available up to now display a low 
   deuterium abundance, D/H~= (0.85~$\pm$ 0.09) $\times$ 10$^{-5}$, while in 
   the intermediate range [$d \sim$100--500 pc; log\,$N$(H\,{\small I}) = 
   19.2--20.7] the D/H ratio is found to vary from $\sim$~0.5~$\times$ 
   10$^{-5}$ to $\sim$~2.2~$\times$~10$^{-5}$ (see, e.g., Linsky et al. 2005). 
   This behaviour is interpreted as due to deuterium depletion on to dust 
   grains (Linsky et al. 2005; see also Prochaska et al. 2005; Oliveira et al. 
   2006), following an original idea by Jura (1982). According to this 
   picture, the most representative value for the total (gas plus dust) D/H 
   ratio within 1 kpc of the Sun would be (D/H)$_{\rmn{LISM}} \ge$ 
   (2.19~$\pm$~0.27)~$\times$~10$^{-5}$ (Linsky et al. 2005). On the other 
   hand, on the basis of D/O and O/H measurements, H\'ebrard \& Moos (2003) 
   suggest that the low D/H value at `large' distances truly reflects the 
   present-epoch D/H. In their scenario, (D/H)$_{\rmn{LISM}}$ should be lower 
   than 1 ~$\times$~10$^{-5}$.

   Solar system observations of $^3$He permit an indirect determination of the 
   deuterium abundance in the Protosolar Cloud (PSC; Geiss \& Reeves 1972). 
   Such an estimate of the deuterium abundance 4.5 Gyr ago, 
   (D/H)$_{\rmn{PSC}}$~= (2.1 $\pm$ 0.5) $\times$ 10$^{-5}$ (Geiss \& 
   Gloeckler 1998), is remarkably similar to both the primordial and the local 
   D/H values, if \emph{the highest} D/H values measured by \emph{FUSE} 
   represent the actual abundance of deuterium in the vicinity of the Sun. It 
   is worth noticing that measures of deuterium in the atmosphere of Jupiter 
   using the Galileo Probe Mass Spectrometer give a similar value, 
   (D/H)$_{\rmn{PSC}}$~= (2.6 $\pm$ 0.7) $\times$ 10$^{-5}$ (Mahaffy et al. 
   1998).

   Finally, detection of deuterium in a molecular cloud at 10 parsecs from the 
   Galactic Centre indicates that (D/H)$_{\rmn{GC}}$~= 
   (1.7~$\pm$~0.3)~$\times$~10$^{-6}$ (Lubowich et al. 2000), five orders of 
   magnitude larger than the predictions of GCE models with no continuous 
   source of deuterium in the bulge (Matteucci, Romano \& Molaro 1999). This 
   discrepancy is suggestive of some replenishment process, likely infall of 
   gas of cosmological composition (Lubowich et al. 2000; Audouze et al. 1976).

   \section{Galactic evolution of deuterium}
   \label{sec.dev}

   Once the primordial abundance of deuterium is settled thanks to (either 
   direct or indirect) observations, the knowledge of the present-day Galactic 
   deuterium abundance is needed to establish the degree of astration suffered 
   by deuterium in the Milky Way. This in turn has profound implications for 
   our understanding of the mechanisms of the formation and evolution of the 
   Galaxy.

   \subsection{The solar neighbourhood}
   \label{subsec:sn}

%
   \begin{figure*}
   \hspace{.25cm}
   \psfig{figure=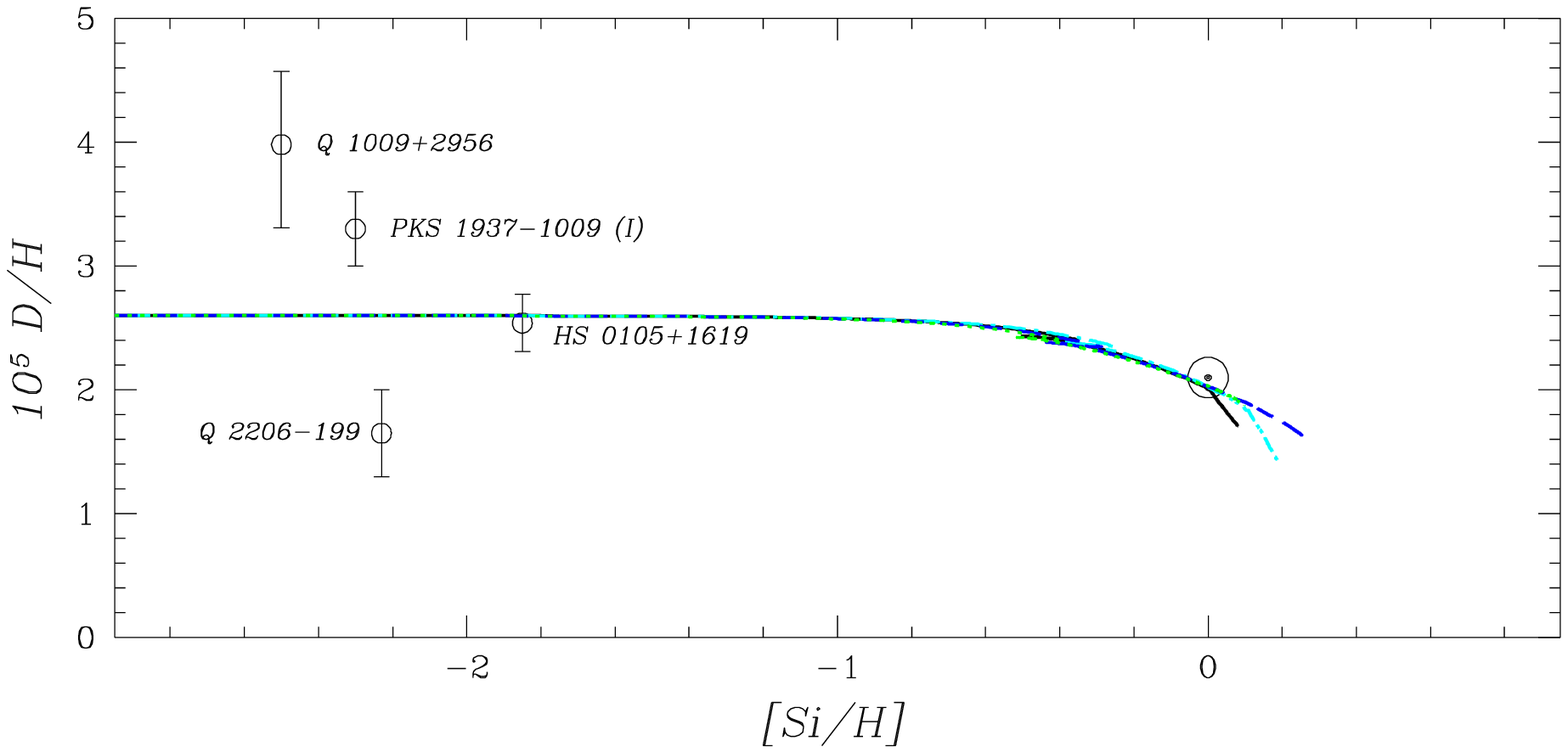,width=12cm}
   \end{figure*}
   \begin{figure*}
   \hspace{.25cm}
   \psfig{figure=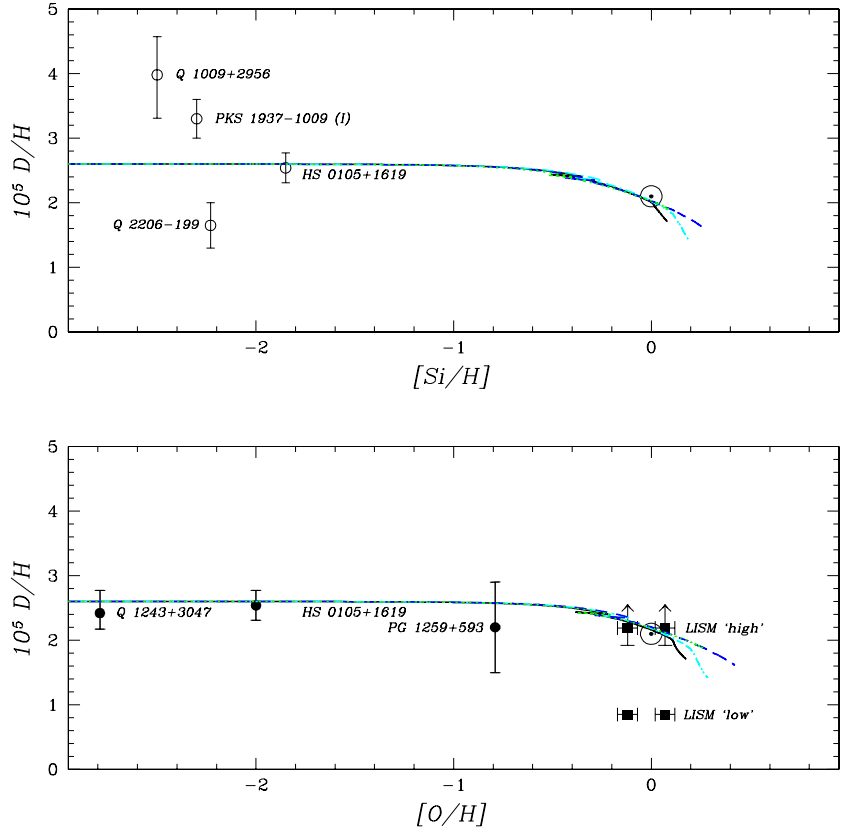,width=12cm}
      \caption{ Theory versus observations. The abundance of deuterium 
	        measured in QSO absorbers (circles) is plotted against 
		metallicity (either [Si/H] -- upper panel, open circles; or 
		[O/H] -- lower panel, filled circles). Only the systems with 
		firm D detections are considered (see Table~\ref{tab.qso} for 
		references and Section~\ref{sec.dat} for discussion). Also 
		shown are the deuterium abundance in the PSC (Geiss \& 
		Gloeckler 1998; upper and lower panels; Sun symbol) and those 
		representative of the true D/H in the LISM (lower panel, 
		squares), according to either Linsky et al. (2005; `high' 
		value) or H\'ebrard \& Moos (2003; `low' value). The values of 
		O/H in the LISM are those measured by Oliveira et al. (2005) 
		for the neutral ISM inside the LB, log(O/H)+12 = 8.54 $\pm$ 
		0.02, and Esteban et al. (2004) for Orion (ionized gas plus 
		dust), log(O/H)+12 = 8.73 $\pm$ 0.03. Notice that the Oliveira 
		et al. measurement refers to the gas phase only, and thus 
		misses $\sim$25 per cent of the total oxygen due to depletion 
		on to dust grains. Theoretical predictions (upper panel: D/H 
		versus [Si/H]; lower panel: D/H versus [O/H]) refer to models 
		for the solar vicinity computed with different prescriptions 
		about the stellar lifetimes and IMF: Scalo's (1986) IMF and 
		Maeder \& Meynet's (1989) stellar lifetimes (solid line); 
		Kroupa et al.'s (1993) IMF and Maeder \& Meynet's (1989) 
		stellar lifetimes (dot-dashed line); Scalo's (1986) IMF and 
		Schaller et al.'s (1992) stellar lifetimes (dotted line); 
		Kroupa et al.'s (1993) IMF and Schaller et al.'s (1992) 
		stellar lifetimes (dashed line). See Romano et al. (2005a) for 
		details about the models. All ratios are normalized to the 
		solar abundances of Asplund et al. (2005).
              }
         \label{fig.mod}
   \end{figure*}
%

   If (D/H)$_{\mathrm{p}}$~= (2.6~$\pm$ 0.3)~$\times$ 10$^{-5}$, and 
   (D/H)$_{\rmn{LISM}} \ge$ (2.19~$\pm$ 0.27)~$\times$ 10$^{-5}$ according to 
   Linsky et al. (2005), then a low astration factor, $f \le$ 1.2~$\pm$~0.3, 
   is required. On the other hand, if (D/H)$_{\rmn{LISM}}$~= 
   (0.85~$\pm$~0.09)~$\times$~10$^{-5}$, as argued by H\'ebrard \& Moos 
   (2003), then $f \simeq$ 3.1~$\pm$~0.7. Here $f$ is the astration factor 
   by number,
   \begin{equation}
     f = \frac{{\rmn{(D/H)_p}}}{{\rmn{(D/H)_{LISM}}}}.
   \label{eq.fnu}
   \end{equation}
   In previous work (Romano et al. 2003), we showed that the astration factor 
   $f \sim$ 1.5 which is required in order to fulfil all the observational 
   constraints available for the Milky Way, well reproduces also the PSC and 
   LISM deuterium abundances, provided that (D/H)$_{\mathrm{LISM}}$~= 
   (1.50~$\pm$~0.10)~$\times$~10$^{-5}$\footnote{Often quoted in GCE works are 
   the astration factors by mass, $F = 
   X_{\rmn{D,\,p}}/X_{{\rmn{D,\,}}t_{{\rmn{now}}}}$, where $X_{\rmn{D,\,p}}$ 
   and $X_{{\rmn{D,\,}}t_{{\rmn{now}}}}$ are the abundances of deuterium by 
   mass at $t$ = 0 and $t$ = $t_{{\rmn{now}}}$ = 13.7 Gyr, respectively. $F$ 
   is larger than $f$ by a few per cent, because of hydrogen burning into 
   helium and heavier species in the course of the Galaxy's evolution.}. Such 
   a low astration factor is due to the combination of a moderate star 
   formation and a continuous infall of external gas during the whole Galactic 
   disc evolution. Noticeably, this result is almost \emph{independent of the 
   specific GCE code used} (either that developed by Tosi 1988a,b; or that 
   developed by Chiappini et al. 1997, 2002). Higher astration factors ($f 
   \sim$ 2--3) were suggested in the past (e.g. Galli et al. 1995; Prantzos 
   1996; Fields 1996; but see also Chiappini et al. 1997; Tosi et al. 1998). 
   The low one reported here is mainly due to the adoption of an updated 
   stellar metallicity distribution, which needs more infall in order to be 
   reproduced. Having that much infall, it comes out really hard to get larger 
   astration factors. As stressed above, in Romano et al. (2003) we adopted 
   the mean D/H value measured by Linsky (1998) in the LB, 
   (D/H)$_{\rmn{LB}}$~= (1.50~$\pm$~0.10)~$\times$~10$^{-5}$, as 
   representative of the deuterium abundance in the solar vicinity at the 
   present time. This assumption seems now to be no longer valid in the light 
   of recent, contrasting interpretations of the dispersion in the local data 
   (see discussions in Section~\ref{sec.dat}). Therefore, in this section we 
   recompute the evolution of D/H.

   First, we adopt various prescriptions for the stellar lifetimes and initial 
   mass function (IMF) in a successful model for the chemical evolution of the 
   Milky Way. Details about the model can be found in Chiappini et al. (2002) 
   and Romano et al. (2005a), where results relevant to several chemical 
   species are discussed. Notice that here we consider only those (IMF, 
   stellar lifetimes) combinations which proved to guarantee a good fit to 
   all the observational constraints available for the Milky Way (see Romano 
   et al. 2005a). The aim is to associate a `theoretical error' to our 
   estimate of $f$.

   In Fig.~\ref{fig.mod} we show model predictions for D/H versus [Si/H] 
   (upper panel) and D/H versus [O/H] (lower panel) in the solar neighbourhood 
   obtained with:
   \begin{enumerate}
     \item the Scalo (1986) IMF and the Maeder \& Meynet (1989) stellar 
           lifetimes (solid line); 
     \item the Kroupa et al. (1993) IMF and the Maeder \& Meynet (1989) 
           stellar lifetimes (dot-dashed line); 
     \item the Scalo (1986) IMF and the Schaller et al. (1992) stellar 
           lifetimes (dotted line);
     \item the Kroupa et al. (1993) IMF and the Schaller et al. (1992) stellar 
           lifetimes (dashed line).
   \end{enumerate}
%
   \begin{table}
   \caption[]{ Theoretical astration factors for deuterium. Listed are the 
               astration factors by number, $f$, suitable for comparison with 
	       the observations. Different astration factors are predicted 
	       when assuming differing formulations for the stellar lifetimes 
	       and IMF in the GCE code.
             }
   \begin{center}
   \begin{tabular}{l l c}
   \hline
   IMF & Stellar lifetimes & Astration factors\\
   \hline
   Scalo (1986)         & Maeder \& Meynet (1989) & 1.52\\
   Kroupa et al. (1993) & Maeder \& Meynet (1989) & 1.83\\
   Scalo (1986)         & Schaller et al. (1992)  & 1.39\\
   Kroupa et al. (1993) & Schaller et al. (1992)  & 1.61\\
   \hline
   \end{tabular}
   \end{center}
   \label{tab.ast}
   \end{table}
%
   Notice that the dotted lines lay on the dashed ones, but they end up with a 
   larger D/H at present, D/H~= 1.9~$\times$~10$^{-5}$ rather than D/H~= 
   1.6~$\times$~10$^{-5}$. Indeed, different D/H and O/H ratios are reached at 
   late times according to the different IMFs and stellar lifetimes adopted. 
   In particular, the higher the mass of massive stars (responsible for a 
   quick recycling of the gas) and the lower the mass of very-low mass objects 
   (which just lock-up material, from the point of view of GCE), the lower 
   (higher) the D/H (O/H) ratio predicted at present. The fraction of stars 
   with initial mass $m \ge$ 8~M$_\odot$ ($m \le$ 1~M$_\odot$) is larger 
   (smaller) for the Kroupa et al. (1993) IMF than for the Scalo (1986) one 
   (see Romano et al. 2005b, their figure~1). This explains the different 
   behaviours predicted by the models shown in Fig.~\ref{fig.mod}. A 
   primordial abundance of (D/H)$_{\rmn{p}}$~= 2.6~$\times$~10$^{-5}$ is 
   assumed for all the models, i.e. the weighted mean of the most reliable QSO 
   absorber data available in the literature (see Table~\ref{tab.qso}). This 
   value agrees well with that inferred from CMB anisotropy measurements and 
   the standard BBN theory. Also shown in Fig.~\ref{fig.mod} are D/H 
   measurements from QSO absorber spectra (circles; see Section~\ref{sec.dat} 
   for references), the PSC deuterium abundance after Geiss \& Gloeckler 
   (1998; Sun symbol), and the representative (D/H)$_{\rmn{LISM}}$ value 
   according to either Linsky et al. (2005; `high' value) or H\'ebrard \& Moos 
   (2003; `low' value). Two [O/H] values are displayed for the LISM, namely 
   that measured by Oliveira et al. (2005) for the neutral medium inside the 
   LB, log(O/H)+12~= 8.54~$\pm$~0.02, and that given by Esteban et al. (2004) 
   for Orion, log(O/H)+12~= 8.73~$\pm$~0.03, which represents the oxygen 
   content of the ionized matter and dust. Notice that part of the local 
   oxygen is depleted on to dust grains ($\sim$25 per cent; see Oliveira et 
   al. 2005; Meyer, Jura \& Cardelli 1998; Andr\'e et al. 2003; Cartledge et 
   al. 2004), which brings the two measurements into agreement. All the ratios 
   are normalized to the solar oxygen and silicon abundances from Asplund et 
   al. (2005). An apparently striking feature of all the models is that only 
   minor deuterium destruction is predicted as [O/H] varies on more than two 
   orders of magnitude, increasing from [O/H] $\le -$3 dex up to [O/H] $\sim$ 
   $-$0.5 dex. The deuterium abundance is sensibly reduced afterwards. This 
   behaviour is easily explained by the large oxygen yield from the first Type 
   II supernovae, which leads to a prompt oxygen enrichment of the ISM only a 
   few million years after the onset of the star formation. Indeed, we find 
   [O/H]~$> -$3.0 dex already 4 million years after the onset of the star 
   formation in the Galaxy, which raises to [O/H]~= $-$0.5 dex in the 
   following $\sim$300 million years. In the meantime, continuing infall of 
   gas of primordial chemical composition helps to keep the D abundance near 
   its primordial value -- infall of gas is particularly strong during the 
   early Galaxy evolution, so that the consumed D is being replenished fast by 
   infall at those earlier times. Later on, when the contribution from low- 
   and intermediate-mass stars (which are not net oxygen producers) to the 
   chemical evolution of the Galaxy becomes increasingly important and the 
   infall rate is reduced, a steeper behaviour on the D/H versus O/H plot is 
   observed.

%
   \begin{figure}
   \psfig{figure=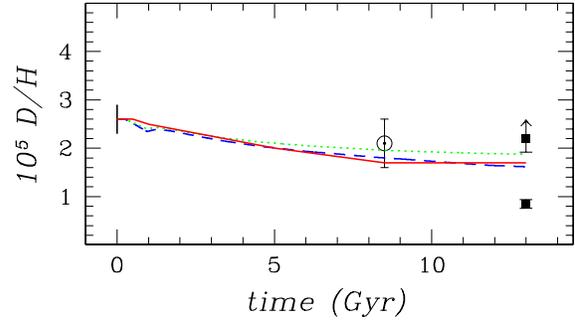,width=\columnwidth}
      \caption{ Evolution of deuterium with time in the solar neighbourhood, 
	        as predicted by two distinct codes for the chemical evolution 
		of the Milky Way. The codes assume different star formation 
		and infall laws, but the same IMF (Kroupa et al. 1993) and 
		stellar lifetimes (Schaller et al. 1992 -- solid and dashed 
		lines; see text for details). Also shown are the predictions 
		from a code assuming Scalo's (1986) IMF and Schaller et al.'s 
		(1992) stellar lifetimes (dotted line). Data: vertical bar at 
		$t$ = 0: mean (D/H)$_{\rmn{p}}$ from QSO absorber observations 
		(see Section~\ref{sec.dat} for references); Sun symbol: PSC D 
		abundance from Geiss \& Gloeckler (1998); squares: fiducial 
		`true' (D/H)$_{\rmn{LISM}}$ value after Linsky et al. (2005; 
		`high' value) and H\'ebrard \& Moos (2003; `low' value).
              }
         \label{fig.com}
   \end{figure}
%

   In Table~\ref{tab.ast} the astration factors by number for all the models 
   displayed in Fig.~\ref{fig.mod} are listed and compared to each other. All 
   the models well reproduce the solar abundances of oxygen and silicon at 
   $t = t_\odot = 9.2$ Gyr. A model in which the Scalo (1986) IMF and the 
   Schaller et al. (1992) stellar lifetimes are adopted is able to destroy D 
   at a level consistent with that suggested by Linsky et al. (2005). None of 
   our models is able to predict astration factors larger than $f \simeq$ 2. 
   This leads us to favour the dust depletion scenario.

%
   \begin{figure}
   \psfig{figure=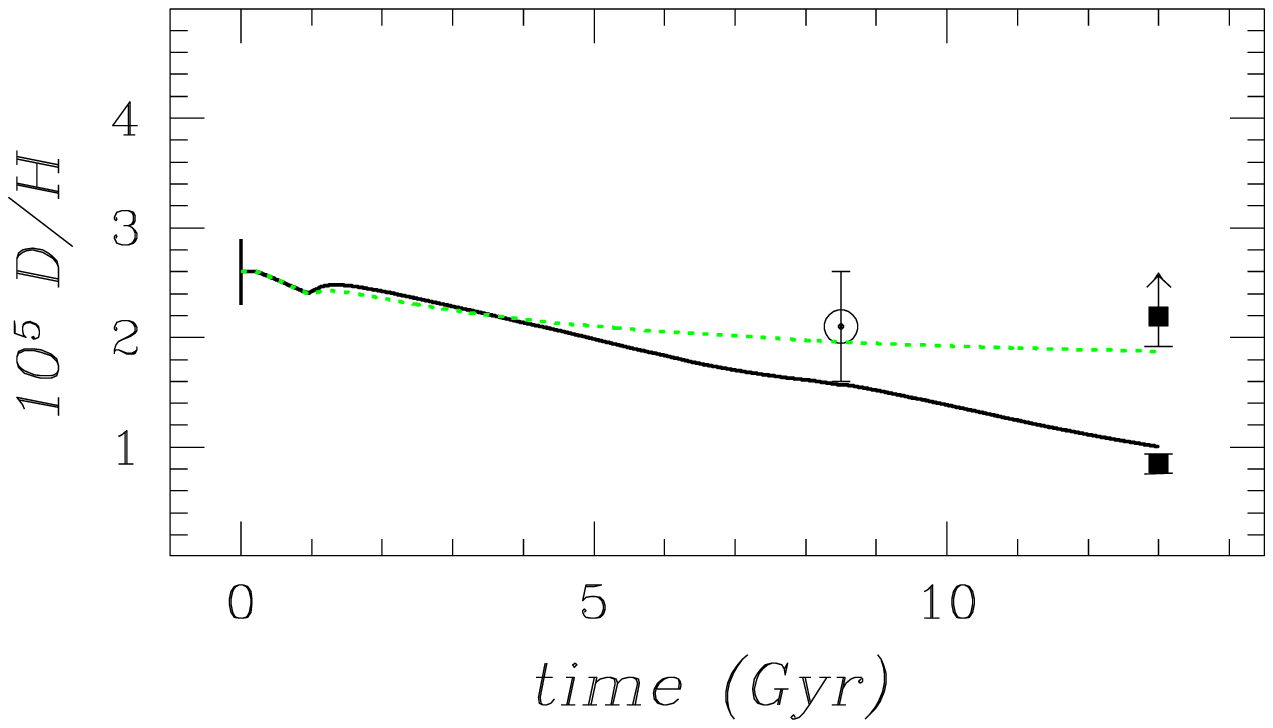,width=\columnwidth}
   \psfig{figure=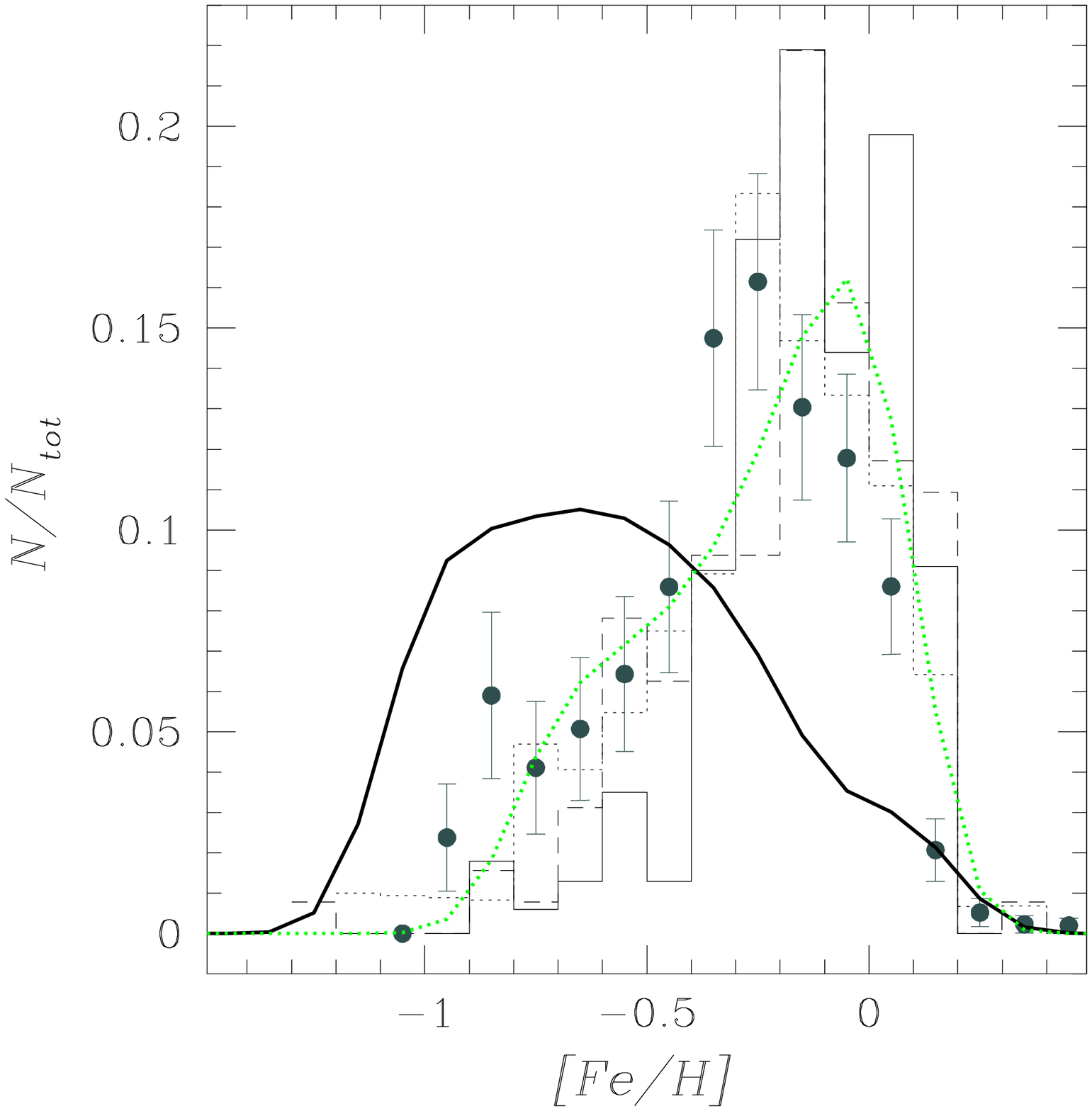,width=\columnwidth}
      \caption{ Models aimed at reproducing the lowest (D/H)$_{\rmn{LISM}}$ 
	        values disagree with other important observational 
		constraints. For instance, the model displayed in this figure 
		(solid lines) assumes a very short time scale for thin-disc 
		formation in the solar neighbourhood ($\tau_{\rmn{D}}$ = 1.5 
		Gyr), at variance with common assumptions (see Chiappini et 
		al. 1997). Hence, while a lower (D/H)$_{\rmn{LISM}}$ can be 
		attained (upper panel), the G-dwarf metallicity distribution 
		of solar neighbourhood stars can not be reproduced any more 
		(lower panel). Also shown for comparison are the predictions 
		of a successful model ($\tau_{\rmn{D}}$ = 7 Gyr; dotted 
		lines), which well reproduces the shape of the observed 
		G-dwarf metallicity distribution. This model leads to a higher 
		(D/H)$_{\rmn{LISM}}$ value. The observed G-dwarf metallicity 
		distributions (histograms in the lower panel) are from Wyse \& 
		Gilmore (1995; dashed histogram), Rocha-Pinto \& Maciel (1996; 
		dotted histogram) and J\o rgensen (2000; solid histogram). 
		Also shown is the K-dwarf metallicity distribution by Kotoneva 
		et al. (2002; big dots). The theoretical distributions are 
		convolved with a Gaussian ($\sigma$ = 0.1) to account for the 
		observational and intrinsic scatter.
              }
         \label{fig.bad}
   \end{figure}
%
   
   Deuterium astration in galaxies is expected to be strongly dependent on the 
   assumptions for the two competing processes of star formation and gas 
   infall. In Fig.~\ref{fig.com} we compare the results obtained for D/H 
   versus time by two GCE codes, the one of Chiappini et al. (2002; dashed 
   line) and that by Tosi (1988a,b; solid line), once adopting the same 
   prescriptions on the stellar lifetimes (i.e. Schaller et al. 1992) and the 
   stellar IMF (Kroupa et al. 1993). The two models assume different 
   prescriptions for the star formation and infall rates, yet the predicted 
   deuterium evolution is almost the same. This is because what really matters 
   is \emph{the interplay} between subtraction and replenishment of gas at 
   each time, and this interconnection is severely constrained by a number of 
   independent observables (see, e.g., Matteucci 2004). According to 
   Fig.~\ref{fig.com}, only a mild deuterium destruction is predicted in the 
   last 4.5 Gyr. This is consistent with the mild Galactic evolution at late 
   times suggested by the small increase of the overall metallicity of the gas 
   from the time of Sun formation up to now (Esteban et al. 2004, and 
   references therein). Also the observed G- and K-dwarf metallicity 
   distributions clearly show that the majority of the stars in the solar 
   neighbourhood formed at low metallicities, [Fe/H] $\le$ 0, thus pointing 
   again to a low star formation activity (and, hence, less evolution) at late 
   times. Also shown in Fig.~\ref{fig.com} is the Chiappini et al. (2002) 
   model with Scalo's (1986) IMF and Schaller et al.'s (1992) stellar 
   lifetimes, which best reproduces both the solar and local `high' D/H values 
   (dotted line). It is worth emphasizing that with these prescriptions the 
   model also well reproduces all the other observational constraints (Romano 
   et al. 2005a).

   In general, our models predict (D/H)$_{\rmn{LISM}}$ values in the range 
   1.4--2.0 $\times$ 10$^{-5}$. We need to enhance the ratio of the star 
   formation to gas infall efficiencies if a larger fraction of the initial 
   deuterium has to be destroyed. But can a more efficient star formation 
   and/or less effective infall of unprocessed gas be accommodated within 
   successful GCE models? It is immediately seen that the almost constant 
   metal abundance of the ISM from the time of Sun's formation up to now 
   (Esteban et al. 2004) sharply contrasts with rapid star formation and/or 
   absence of external sources of unenriched gas. In Fig.~\ref{fig.bad}, we 
   show the outputs of a model especially designed to produce a significant 
   drop in deuterium abundance from its PSC value to the local one (for 
   further modeling and discussions see Tosi et al. 1998). This model, 
   assuming $\tau_{\rmn{D}}$ = 1.5 Gyr rather than 7 Gyr for the thin-disc 
   formation time scale, fits the solar and local `low' D/H values 
   (Fig.~\ref{fig.bad}, upper panel, solid line), as well as other 
   observational constraints (such as for instance the current gaseous and 
   stellar mass surface densities at the solar position). However, it is 
   barely consistent with the metal abundances of the Sun, it predits a sharp 
   increase of the metal abundance of the ISM in the last 4.5~Gyr (at variance 
   with the observations), and it can not account for the metallicity 
   distribution of solar neighbourhood G- and K-dwarf stars 
   (Fig.~\ref{fig.bad}, lower panel, solid line). The metallicity distribution 
   of long-lived stars in the solar neighbourhood is a fundamental constraint 
   for GCE models, which mainly constrains the infall time scale and strength 
   (see, e.g., Chiappini et al. 1997; Matteucci 2004). In Fig.~\ref{fig.bad}, 
   lower panel, we show as a dashed, dotted, and solid histogram the G-dwarf 
   metallicity distributions observed by Wyse \& Gilmore (1995), Rocha-Pinto 
   \& Maciel (1996) and J\o rgensen (2000), respectively. Also shown is the 
   K-dwarf metallicity distribution by Kotoneva et al. (2002; big dots). K 
   dwarfs have lifetimes older than the present age of the Galactic disc, and 
   are thus ideal stars for investigating the chemical evolution of the disc. 
   G dwarf stars instead are sufficiently massive that some of them have begun 
   to evolve away from the main sequence, so that corrections must be taken 
   into account when determining their space densities and metallicity 
   (Kotoneva et al. 2002). These corrections are at least partly responsible 
   for the non-negligible differences among different observational G-dwarf 
   metallicity distributions (in particular, the low-metallicity tail of the 
   distribution and the exact position of the peak may vary significantly 
   according to different authors). In conclusion, in the framework of our 
   models we can not explain at the same time both the lowest observed 
   (D/H)$_{\rmn{LISM}}$ values and the observed G- and K-dwarf metallicity 
   distributions.

   \subsection{Present-day abundance gradients}
   \label{subsec:gr}

%
   \begin{figure*}
   \hspace{.25cm}
   \psfig{figure=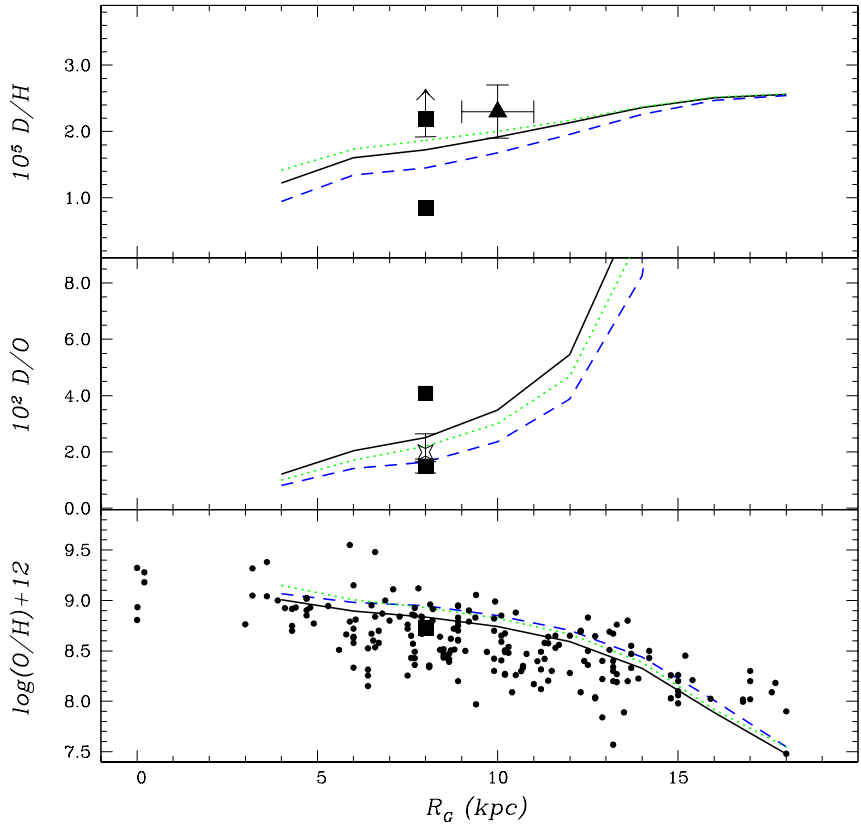,width=12cm}
      \caption{ Present-day distributions of D/H (upper panel), D/O (middle 
                panel) and log(O/H)+12 (lower panel) along the disc of the 
		Galaxy. Predictions from models assuming either the Scalo 
		(1986) IMF and the Maeder \& Meynet (1989) stellar lifetimes 
		(solid line), or the Kroupa et al. (1993) IMF and the Maeder 
		\& Meynet (1989) stellar lifetimes (dashed line), or the Scalo 
		(1986) IMF and the Schaller et al. (1992) stellar lifetimes 
		(dotted line) are shown together with the relevant 
		observations. Upper panel, filled squares: representative D/H 
		value in the LISM according to Linsky et al. (2005; `high' 
		value) and H\'ebrard \& Moos (2003; `low' value); filled 
		triangle: deuterium measurement from the 327 MHz D line in the 
		outer disc (Rogers et al. 2005). Middle panel, filled squares: 
		representative D/O value in the LISM according to H\'ebrard \& 
		Moos (2003; `low' value) and Linsky et al. (2005; `high' 
		value -- in this case the oxygen abundance measured by Esteban 
		et al. 2004 for Orion has been taken as representative of the 
		total oxygen content of the LISM). Also shown (big star) is 
		the D/O ratio towards LSE\,44 measured by Friedman et al. 
		(2006; see discussion in the text). Lower panel, small filled 
		circles: compilation of log(O/H)+12 versus $R_{\rmn{G}}$ data 
		from Chiappini et al. (2001); filled square: oxygen abundance 
		in Orion from Esteban et al. (2004).
              }
         \label{fig.grad}
   \end{figure*}
%

%
   \begin{figure*}
   \hspace{.25cm}
   \psfig{figure=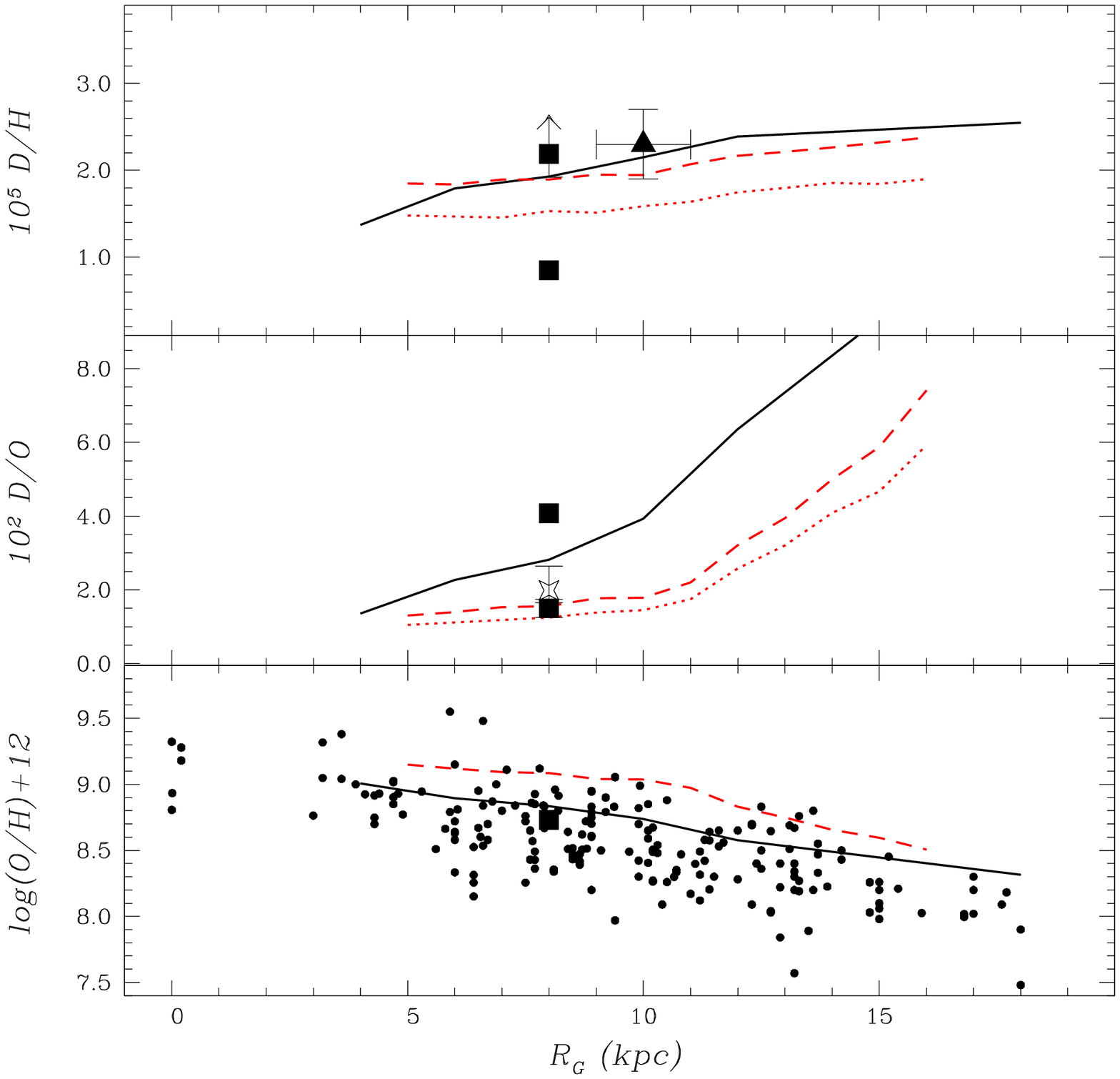,width=12cm}
      \caption{ Same as Fig.~\ref{fig.grad}, but for Model C of Chiappini et 
	        al. (2001; solid lines) and for the model of Tosi et al. (in 
		preparation), which is an updated version of the model of Tosi 
		(1988a,b) extended to the whole disc (dashed and dotted 
		lines). Here (D/H)$_{\rmn{p}}$~= 2.9~$\times$~10$^{-5}$ (solid 
		and dashed lines) or 2.3~$\times$~10$^{-5}$ (dotted lines), 
		and not 2.6~$\times$~10$^{-5}$.
              }
         \label{fig.grad2}
   \end{figure*}
%

   In Fig.~\ref{fig.grad} we display the Galactic gradients of D/H, D/O, and 
   O/H at the present time predicted by adopting the prescriptions of Model A 
   of Chiappini et al. (2001) as far as the formation and evolution of the 
   Galactic disc are concerned, and by varying the prescriptions for the IMF 
   and the stellar lifetimes (solid line: Scalo's 1986 IMF and Maeder \& 
   Meynet's 1989 stellar lifetimes; dashed line: Kroupa et al.'s 1993 IMF and 
   Maeder \& Meynet's 1989 stellar lifetimes; dotted line: Scalo's 1986 IMF 
   and Schaller et al.'s 1992 stellar lifetimes -- see Romano et al. 2005a). 
   A value of (D/H)$_{\rmn{p}}$~= 2.6~$\times$~10$^{-5}$ is assumed for all 
   the models. Also shown are the relevant observations. The filled squares in 
   the upper panel of Fig.~\ref{fig.grad} stand for the representative D/H 
   value in the LISM according to either Linsky et al. (2005; `high' value) or 
   H\'ebrard \& Moos (2003; `low' value). The filled triangle in the same 
   panel refers to the measurement by Rogers et al. (2005) for the outer 
   Galactic disk. The filled squares in the middle panel of 
   Fig.~\ref{fig.grad} stand for the representative D/O ratio according to 
   either H\'ebrard \& Moos (2003; lower value) or Linsky et al. (2005; higher 
   value). To obtain (D/O)$_{\rmn{LISM}}$ in this latter case, we divided 
   (D/H)$_{\rmn{LISM}}$ from Linsky et al. (2005) by the oxygen abundance of 
   Orion (Esteban et al. 2004). At first glance, it seems that GCE models 
   explain the lowest D/O value better than the highest one, while the 
   opposite is true for D/H (upper panel). However, one should recall that the 
   current oxygen abundance displays a non-negligible spread. Indeed, several 
   indicators -- H\,{\small II} regions, B-type stars, the neutral medium -- 
   show that the local oxygen abundance varies in the range 8.3~$\le$ 
   log(O/H)+12~$\le$ 9.0 (Fig.~\ref{fig.grad}, lower panel, and Oliveira et 
   al. 2006, their table~12). So, in Fig.~\ref{fig.grad}, middle panel, we 
   show also the D/O ratio recently measured by Friedman et al. (2006) for 
   the line of sight towards LSE\,44. Along this line of sight, 
   (D/H)$_{\rmn{LSE \ 44}}$~= 2.24$^{+0.70}_{-0.67}$~$\times$~10$^{-5}$, 
   consistent with the lower limit suggested by Linsky et al. (2005), and 
   (O/H)$_{\rmn{LSE \ 44}}$~= 11.3$^{+4.8}_{-3.6}$~$\times$~10$^{-4}$ (that is 
   the highest O/H value measured locally by \emph{FUSE}). Now, within the 
   errors, the model adopting the Scalo (1986) IMF and the Schaller et al. 
   (1992) stellar lifetimes is in satisfactory agreement with both the 
   observed `high' (D/H)$_{\rmn{LISM}}$ and `low' (D/O)$_{\rmn{LISM}}$, 
   provided that (D/O)$_{\rmn{LISM}}$ is `low' as a consequence of both a high 
   deuterium and a high oxygen abundance in the LISM.

   It is seen that the shape of the gradients slightly changes when changing 
   the prescriptions about the IMF and the stellar lifetimes. In the 
   Galactocentric distance range $R_{\rmn{G}}$ = 4--12 kpc the theoretical 
   gradients of log(D/H), log(D/O), and log(O/H)+12 are 0.03~$\pm$~0.01 
   dex~kpc$^{-1}$, 0.08 dex~kpc$^{-1}$, and $-$0.055~$\pm$~0.005 
   dex~kpc$^{-1}$, respectively. In the outermost regions of the disc the 
   gradient of D/H flattens, while those of D/O and O/H steepen, owing to the 
   lower and lower amount of gas processed by stars outward in the disc.

   Before concluding that, according to GCE model predictions, one shall 
   observe D/H ratios close to the primordial value at the largest radii in 
   spiral galaxies, it is worth reminding that the model predictions depend on 
   a number of uncertain free parameters. Model A of Chiappini et al. (2001) 
   produces a fairly steep O/H gradient at the largest radii 
   (Fig.~\ref{fig.grad}, bottom panel), while the majority of the data suggest 
   a flatter behaviour. A flatter O/H gradient in the outer disc is easily 
   obtained by suppressing the gas density threshold which regulates the star 
   formation process during the preceding halo phase. The solid lines in 
   Fig.~\ref{fig.grad2} stand for Chiappini et al.'s (2001) Model C results. 
   This model is analogous to Model A (Fig.~\ref{fig.grad}, solid lines), 
   except that it allows star formation in the halo to go on even when the gas 
   density drops below the threshold value ($\sim$~4 M$_\odot$~pc$^{-2}$). The 
   expected D/H ratio is now $\sim$~2.5~$\times$~10$^{-5}$ at the outermost 
   radii, i.e. below the adopted primordial value [(D/H)$_{\rmn{p}}$~= 
   2.9~$\times$~10$^{-5}$ for this model]. Also shown in Fig.~\ref{fig.grad2} 
   are the predictions from the model of Tosi et al. (in preparation), which 
   is an updated version of the model by Tosi (1988a,b) adopting the Kroupa et 
   al. (1993) IMF and the Schaller et al. (1992) stellar lifetimes, and 
   extending to the whole disc [dashed lines: (D/H)$_{\rmn{p}}$~= 
   2.9~$\times$~10$^{-5}$; dotted lines: (D/H)$_{\rmn{p}}$~= 
   2.3~$\times$~10$^{-5}$]. This model tends to predict higher oxygen 
   abundances, hence lower D/O ratios, across the disc.

   We conclude that, in principle, accurate measurements of both D/H and O/H 
   in the outermost regions of spiral galaxies can allow us to discriminate 
   among existing scenarios of galaxy formation and evolution. In fact, only 
   minor uncertainties related to the adopted stellar yields affect the 
   interpretation of the data in this case, owing to the well-known 
   nucleosynthetic origin of both species. Hence, observational efforts in 
   this direction are highly desirable.

   In summary, notice that all the GCE models discussed above favour minor 
   deuterium variations [$\Delta$log(D/H)/$\Delta R_{\rmn{G}} \simeq$ 0.02 dex 
   kpc$^{-1}$] in the Galactocentric distance range sampled by current 
   observations ($R_{\rmn{G}} \simeq$ 8--11 kpc), hence favouring the highest 
   (D/H)$_{\rmn{LIMS}}$ determinations as representative of the local 
   deuterium abundance.

   \section{Discussion and conclusions}
   \label{sec.fin}

   In general, observations of deuterium abundances provide \emph{lower 
   bounds} on its primordial abundance. However, at sufficiently high 
   redshifts and low metallicities, one expects the primordial abundance of 
   deuterium to reveal itself as a plateau. Instead, statistically significant 
   scatter of the D/H measurements has been found in several high-redshift QSO 
   absorbers. Before claiming that a dispersion around the mean value is 
   actually present, one has to carefully check possible systematics affecting 
   the abundance derivation. It has been suggested that in the absorbers with 
   the lowest H\,{\small I} column densities, interlopers might indeed 
   contribute to the inferred D\,{\small I} column densities, while in those 
   with the highest H\,{\small I} column densities, interlopers might affect 
   the wings of the H\,{\small I} lines (Steigman 2004, 2006, and references 
   therein). Yet, it is worth noticing that the placement of the continuum 
   and the fitting of the damping wings are closely related for DLAs: if one 
   gets the continuum wrong, it is unlikely that he could fit the line profile 
   satisfactorily (M. Pettini, private communication). In this paper, we 
   assume (D/H)$_{\rmn{p}}$~= (2.6~$\pm$~0.3)~$\times$~10$^{-5}$, i.e. the 
   weighted mean of D/H measurements towards QSOs (indirect determinations 
   give a consistent value; see discussions in Sections~\ref{sec.int} and 
   \ref{sec.dat}).

   Of more concern is the dispersion of local D/H data. In the LISM, D/H is 
   found to vary from $\sim$~0.5$\times$~10$^{-5}$ to 
   $\sim$~2.2$\times$~10$^{-5}$ (H\'ebrad et al. 2005; Linsky et al. 2005; 
   Williger et al. 2005; Friedman et al. 2006; Oliveira et al. 2006; to quote 
   only the most recent works). Basically, two mechanisms are suggested in the 
   current literature which might explain the observed variation (Pettini 
   2006, and references therein; see also Lemoine et al. 1999; Prochaska et 
   al. 2005): 
   \begin{enumerate}
     \item differing ISM conditions along different sight lines, which 
           determine different degrees of deuterium depletion on to dust 
	   grains;
     \item localized infall of unprocessed gas, which modifies the deuterium 
           abundance in the surveyed region while leaving unchanged those of 
	   more abundant species, such as oxygen.
   \end{enumerate}
   In the first case, one should recover the true local D/H in interstellar 
   clouds which have been accelerated by supernova shocks, since the depletion 
   of refractory elements is much reduced relative to that of the normal 
   quiescent ISM there; in this case, the true local abundance of deuterium 
   would be the highest observed one (Linsky et al. 2005, and references 
   therein). In the second case, the true local abundance of deuterium would 
   be the lowest observed one. In any case, the mean D/H value measured within 
   the Local Bubble should not be considered any more as representative of the 
   actual degree of astration suffered by deuterium in the solar 
   neighbourhood. By assuming that the primordial abundance of deuterium is 
   reasonably well known, one can usefully bind the deuterium astration factor 
   of the solar vicinity. An astration factor (by number) either as low as 
   $f \le 1.2 \pm 0.3$ or a factor of 2--3 higher is permitted by the 
   observations.

   In this paper, we have computed a number of GCE models for the solar 
   vicinity as well as for the whole Galactic disc and conclude that:
   \begin{enumerate}
     \item Only low astration factors (not in excess of $f \simeq 1.8$) are 
           compatible with GCE requirements: our models predict 
	   (D/H)$_{\rmn{LISM}}$~= 1.4--2.0~$\times$~10$^{-5}$ starting from a 
	   primordial value of (D/H)$_{\rmn{p}}$~= 2.6~$\times$~10$^{-5}$. 
	   This is not surprising: in the absence of supernova-driven winds 
	   which efficiently remove the metal-rich ejecta of dying stars from 
	   the Galaxy and/or in the absence of a peculiar IMF, the relatively 
	   high present-day gas content and low metallicity are indicative of 
	   modest astration (and, hence, modest D destruction; see also Tosi 
	   et al. 1998). We emphasize here once again that the low astration 
	   factors are due to the combination of moderate star formation and 
	   continuous infall of gas which are needed in order to reproduce the 
	   available Milky Way data.
     \item Small variations in the predicted D astration factor are produced 
           by changing the prescriptions on the IMF and the stellar lifetimes.
	   In particular, by adopting the Scalo (1986) IMF and the Schaller et 
	   al. (1992) stellar lifetimes we predict a D astration factor which 
	   clearly favours the low one suggested by Linsky et al. (2005). With 
	   this choice for the IMF and the stellar lifetimes, the model also 
	   gives a good fit to many other observational constraints for the 
	   Galaxy (Romano et al. 2005a).
     \item When the model is forced to reproduce the lowest D/H values 
           observed locally, the agreement between model predictions and 
	   relevant observations is lost, for one or more observables (the 
	   most relevant observational constraints being -- in this context -- 
	   the G-dwarf metallicity distribution, the mild increase of the 
	   overall metallicity of the ISM from the time of Sun's formation up 
	   to now, the present-day mass and gas surface densities, the 
	   present-day infall rate; see also Matteucci 2004).
     \item In order to attain the lowest D/H values observed in the LISM, the 
           smooth, gentle decline which nicely accounts for the PSC D 
	   abundance, must turn into a steeper behaviour during the latest 
	   phases of Galaxy's evolution. However, a large gas cycling through 
	   stars in the last $\sim$5 Gyr is unlikely, in the light of the 
	   small increase of the global metal abundance from the time of Sun's 
	   formation up to now suggested by several independent indicators 
	   (Esteban et al. 2004; their table~15 and references therein).
     \item In principle, joint observations of deuterium and oxygen abundances 
           in the outermost regions of galactic discs can shed light on the 
	   mechanisms of spiral galaxy formation and evolution.
   \end{enumerate}
   In conclusion, we favour a scenario in which D/H in the solar neighbourhood 
   declines mildly during Galaxy's evolution, due to the combined effect of 
   moderate star formation and continuous infall of external gas. In this 
   framework, the dispersion in the current local D/H data might be at least 
   partly explained by different degrees of dust depletion along different 
   lines of sight.

   \section*{Acknowledgments}

   DR and MT wish to thank J. Geiss, G. Gloeckler, G. H\'ebrard, J. Linsky, 
   and T. Bania of the LoLa-GE Team for useful conversations at the 
   International Space Science Institute in Bern. The warm hospitality and the 
   financial support at ISSI are gratefully acknowledged. We are also grateful 
   to G. Steigman for stimulating discussions and to W. Moos and M. Pettini 
   for interesting conversations about many observational aspects. We thank J. 
   Linsky and M. Pettini for providing us with a copy of their papers in 
   advance of publication.

\bsp

\label{lastpage}


\begin{thebibliography}{90}

\bibitem[]{}
Andr\'e M. K., Oliveira C. M., Howk J. C., et al., 2003, ApJ, 591, 1000
\bibitem[]{}
Asplund M., Grevesse N., Sauval A. J., 2005, in Barnes T.G. III, Bash F. N., 
   eds, ASP Conf. Ser., Cosmic Abundances as Records of Stellar Evolution and 
   Nucleosynthesis. Astron. Soc. Pac., San Francisco, Vol.~336, p.~25
\bibitem[]{}
Audouze J., Lequeux J., Reeves H., Vigroux L., 1976, ApJ, 208, L51
\bibitem[]{}
Boesgaard A. M., Steigman G., 1985, ARA\&A, 23, 319
\bibitem[]{}
Burles S., Tytler D., 1998a, ApJ, 499, 699
\bibitem[]{}
Burles S., Tytler D., 1998b, ApJ, 507, 732
\bibitem[]{}
Cartledge S. I. B., Lauroesch J. T., Meyer D. M., Sofia U. J., 2004, ApJ, 613, 
   1037
\bibitem[]{}
Charbonnel C., Primas F., 2005, A\&A, 442, 961
\bibitem[]{}
Chiappini C., Matteucci F., Gratton R., 1997, ApJ, 477, 765
\bibitem[]{}
Chiappini C., Matteucci F., Romano D., 2001, ApJ, 554, 1044
\bibitem[]{}
Chiappini C., Renda A., Matteucci F., 2002, A\&A, 395, 789
\bibitem[]{}
Coc A., Vangioni-Flam E., Descouvemont P., Adahchour A., Angulo C., 2004, ApJ, 
   600, 544
\bibitem[]{}
Crighton N. H. M., Webb J. K., Carswell R. F., Lanzetta K. M., 2003, MNRAS, 
   345, 243
\bibitem[]{}
Crighton N. H. M., Webb J. K., Ortiz-Gil A., Fern\'andez-Soto A., 2004, MNRAS, 
   355, 1042
\bibitem[]{}
Epstein R. L., 1977, ApJ, 212, 595
\bibitem[]{}
Epstein R. L., Arnett W. D., Schramm D. N., 1976, ApJS, 31, 111
\bibitem[]{}
Esteban C., Peimbert M., Garc\'\i a-Rojas J., Ruiz M. T., Peimbert A., 
   Rodr\'\i guez M., 2004, MNRAS, 355, 229
\bibitem[]{}
Fields B. D., 1996, ApJ, 456, 478
\bibitem[]{}
Fields B. D., Olive K. A., Silk J., Cass\'e M., Vangioni-Flam E., 2001, ApJ, 
   563, 653
\bibitem[]{}
Friedman S. D., H\'ebrard G., Tripp T. M., Chayer P., Sembach K. R., 2006, 
   ApJ, 638, 847
\bibitem[]{}
Galli D., Palla F., Ferrini F., Penco U., 1995, ApJ, 443, 536
\bibitem[]{}
Geiss J., Gloeckler G., 1998, Space Sci. Rev., 84, 239
\bibitem[]{}
Geiss J., Reeves H., 1972, A\&A, 18, 126
\bibitem[]{}
H\'ebrard G., Moos H. W., 2003, ApJ, 599, 297
\bibitem[]{}
H\'ebrard G., Tripp T. M., Chayer P., et al., 2005, ApJ, 635, 1136
\bibitem[]{}
Jenkins E. B., Tripp T. M., Wo\'zniak P. R., Sofia U. J., Sonneborn G., 
   1999, ApJ, 520, 182
\bibitem[]{}
J\o rgensen B. R., 2000, A\&A, 363, 947
\bibitem[]{}
Jura M., 1982, in Kondo Y., ed., Advances in Ultraviolet Astronomy. 
   Washington, NASA, p.~54
\bibitem[]{}
Kirkman D., Tytler D., Suzuki N., O'Meara J. M., Lubin D., 2003, ApJS, 149, 1
\bibitem[]{}
Kotoneva E., Flynn C., Chiappini C., Matteucci F., 2002, MNRAS, 336, 879
\bibitem[]{}
Kroupa P., Tout C. A., Gilmore G., 1993, MNRAS, 262, 545
\bibitem[]{}
Lemoine M., Audouze J., Ben Jaffel L., et al., 1999, New Astr., 4, 231
\bibitem[]{}
Levshakov S. A., Dessauges-Zavadsky M., D'Odorico S., Molaro P., 2002, ApJ, 
   565, 696
\bibitem[]{}
Linsky J. L., 1998, Space Sci. Rev., 84, 285
\bibitem[]{}
Linsky J. L., Draine B. T., Moos H. W., et al., 2005, BAAS, 37, 1444
\bibitem[]{}
Lubowich D. A., Pasachoff J. M., Balonek T. J., Millar T. J., Tremonti C., 
   Roberts H., Galloway R. P., 2000, Nat, 405, 1025
\bibitem[]{}
Maeder A., Meynet G., 1989, A\&A, 210, 155
\bibitem[]{}
Mahaffy P. R., Donahue T. M., Atreya S. K., Owen T. C., Niemann H. B., 1998, 
   Space Sci. Rev., 84, 251
\bibitem[]{}
Matteucci F., 2004, in McWilliam A., Rauch M., eds, Carnegie Obs. Astroph. 
   Ser., Origin and Evolution of the Elements. Cambridge Univ. Press, 
   Cambridge, p.~87
\bibitem[]{}
Matteucci F., Romano D., Molaro P., 1999, A\&A, 341, 458
\bibitem[]{}
Meyer D. M., Jura M., Cardelli J. A., 1998, ApJ, 493, 222
\bibitem[]{}
Moos H. W., Sembach K. R., Vidal-Madjar A., et al., 2002, ApJS, 140, 3
\bibitem[]{}
Olive K. A., Steigman G., Walker T. P., 2000, Phys. Rep., 333, 389
\bibitem[]{}
Olive K. A., Skillman E. D., 2004, ApJ, 617, 29
\bibitem[]{}
Oliveira C. M., Dupuis J., Chayer P., Moos H. W., 2005, ApJ, 625, 232
\bibitem[]{}
Oliveira C. M., Moos H. W., Chayer P., Kruk J. W., 2006, ApJ in press 
   (astro-ph/0601114)
\bibitem[]{}
O'Meara J. M., Tytler D., Kirkman D., Suzuki N., Prochaska J. X., Lubin D., 
   Wolfe A. M., 2001, ApJ, 552, 718
\bibitem[]{}
Pettini M., 2006, in Sonneborn G., Moos H. W., Andersson B.-G., eds, ASP Conf. 
   Ser., Astrophysics in the Far Ultraviolet. Astron. Soc. Pac., San 
   Francisco, Vol.~348, in press (astro-ph/0601428)
\bibitem[]{}
Pettini M., Bowen D. V., 2001, ApJ, 560, 41
\bibitem[]{}
Prantzos N., 1996, A\&A, 310, 106
\bibitem[]{}
Prantzos N., Ishimaru Y., 2001, A\&A, 376, 751
\bibitem[]{}
Prochaska J. X., Tripp T. M., Howk J. C., 2005, ApJ, 620, L39
\bibitem[]{}
Prodanovi\'c T., Fields B. D., 2003, ApJ, 597, 48
\bibitem[]{}
Rocha-Pinto H. J., Maciel W. J., 1996, MNRAS, 279, 447
\bibitem[]{}
Rogers A. E. E., Dudevoir K. A., Carter J. C., Fanous B. J., Kratzenberg E., 
   Bania T. M., 2005, ApJ, 630, L41
\bibitem[]{}
Rogerson J. B., York D. G., 1973, ApJ, 186, L95
\bibitem[]{}
Romano D., Chiappini C., Matteucci F., Tosi M., 2005a, A\&A, 430, 491
\bibitem[]{}
Romano D., Chiappini C., Matteucci F., Tosi M., 2005b, in Corbelli E., Palla 
   F., Zinnecker H., eds, ASSL, The Initial Mass Function 50 years later. 
   Springer, Dordrecht, Vol.~327, p.~231 (astro-ph/0409474)
\bibitem[]{}
Romano D., Tosi M., Matteucci F., Chiappini C., 2003, MNRAS, 346, 295
\bibitem[]{}
Scalo J. M., 1986, Fund. Cosmic Phys., 11, 1
\bibitem[]{}
Schaller G., Schaerer D., Meynet G., Maeder A., 1992, A\&AS, 96, 269
\bibitem[]{}
Scully S., Cass\'e M., Olive K. A., Vangioni-Flam E., 1997, ApJ, 476, 521
\bibitem[]{}
Sembach K. R., Wakker B. P., Tripp T. M., et al., 2004, ApJS, 150, 387
\bibitem[]{}
Sonneborn G., Tripp T. M., Ferlet R., Jenkins E. B., Sofia U. J., 
   Vidal-Madjar A., Wo\'zniak P. R., 2000, ApJ, 545, 277
\bibitem[]{}
Spergel D. N., Verde L., Peiris H. V., et al., 2003, ApJS, 148, 175
\bibitem[]{}
Steigman G., 1994, MNRAS, 269, L53
\bibitem[]{}
Steigman G., 2004, in Freedman W. L., ed., Carnegie Obs. Astr. Ser., Measuring 
   and Modeling the Universe. Cambridge, Cambridge Univ. Press, p.~169
\bibitem[]{}
Steigman G., 2006, Int. Journal of Modern Phys. E, 15, 1
\bibitem[]{}
Steigman G., Tosi M., 1992, ApJ, 401, 150
\bibitem[]{}
Steigman G., Tosi M., 1995, ApJ, 453, 173
\bibitem[]{}
Tosi M., 1988a, A\&A, 197, 33
\bibitem[]{}
Tosi M., 1988b, A\&A, 197, 47
\bibitem[]{}
Tosi M., 1996, in Leitherer C., Fritze-von-Alvensleben U., Huchra J., eds, ASP 
   Conf. Ser., From Stars to Galaxies: The Impact of Stellar Physics on Galaxy 
   Evolution. Astron. Soc. Pac., San Francisco, Vol.~98, p.~299
\bibitem[]{}
Tosi M., Steigman G., Matteucci F., Chiappini C. 1998, ApJ, 498, 226
\bibitem[]{}
Vangioni-Flam E., Olive K. A., Prantzos N. 1994, ApJ, 427, 618
\bibitem[]{}
Weinberg S., 1977, The first three minutes. A modern view of the origin of the 
   universe. London: Andre Deutsch
\bibitem[]{}
Williger G. M., Oliveira C., H\'ebrard G., Dupuis J., Dreizler S., Moos H. W., 
   2005, ApJ, 625, 210
\bibitem[]{}
Wood B. E., Linsky J. L., H\'ebrard G., Williger G. M., Moos H. W., Blair W. 
   P., 2004, ApJ, 609, 838
\bibitem[]{}
Wyse R. F. G., Gilmore G., 1995, AJ, 110, 2771
\end{thebibliography}
\end{document}